\documentstyle[12pt,epsf]{article}

\def\be{\begin{equation}}
\def\ee{\end{equation}}
\def\ba{\begin{eqnarray}}
\def\ea{\end{eqnarray}}

\def\esp#1{{\rm e}^{#1}}
\def\ii{{\rm i}}
\def\tr{{\rm tr}}

\def\diag{{\rm diag}}

\def\a{\alpha}
\def\adot{{\dot\a}}
\newcommand{\Dslash}{{D \hspace{-7.4pt} \slash}\;}
\begin{document}
\rightline{DFTT-2/99}
\rightline{KUL-TF-99/3}
\rightline{\hfill January 1999}
\vskip 0.8cm
\centerline{\Large \bf 
2D Yang--Mills Theory as a Matrix String Theory\footnote{%
Talk presented by A. D'Adda at the 2nd 
Conference on Quantum Aspects of Gauge  Theories, 
Supersymmetry and Unification, Corfu, Greece, 21-26 September 1998.
Work partially supported by the European Commission
TMR programme ERBFMRX-CT96-0045.}}
\vskip 0.8cm
\centerline{\bf M. Bill\'o${}^{1,2}$, M.\,Caselle${}^2$,
A.\,D'Adda${}^2$ and P. Provero${}^2$}
\vskip 0.5cm
\centerline{${}^1$ \sl Instituut voor Theoretische Fysica, K.U. Leuven,}
\centerline{\sl Celestijnenlaan 200D, B3001 Leuven, Belgium}
\vskip 0.2cm
\centerline{${}^2$ \sl Dipartimento di Fisica
Teorica dell'Universit\`a di Torino and } 
\centerline{\sl I.N.F.N., Sezione di Torino, via P.Giuria 1, I-10125 Torino, 
Italy}
\vskip 0.2cm
\centerline{\tt billo,caselle,dadda,provero@to.infn.it}
\begin{abstract}
Quantization of two--dimensional Yang--Mills theory on a torus in the gauge
where the field strength is diagonal leads to twisted sectors that are
completely analogous to the ones that originate long string states in Matrix
String Theory. If these sectors are taken into account the partition function
is different from the standard one found in the literature and the invariance
of the theory under modular transformations of the torus 
appears to hold in a stronger sense. 
The twisted sectors are in one-to-one correspondence with the coverings
of the torus without branch points, so they define by themselves a string
theory. A possible duality between this string theory and the Gross-Taylor
string is discussed, and the problems that one encounters in generalizing this
approach to interacting strings are pointed out.
This talk is based on a previous paper by the same authors, but it contains
some new results and a better interpretation of the results already obtained. 
\end{abstract}
\section{Matrix String Theory}
This talk is based on  Ref. \cite{us}, but it also includes
some new results and, we hope, a better understanding of the 
results already contained in the original paper.
This work has its origin in an old puzzle that dates back to 1993 
and in some recent development. 
The puzzle consists in the following: while studying the functional 
integral approach to quantization of YM2 on a torus (see Ref. \cite{us93})
some of us noticed that in order to get the correct result for the partition
function on the torus in the gauge where the field strength $F$ is 
diagonal (unitary gauge) we would have been obliged to neglect some
contributions from twisted sectors that seem to 
arise naturally in that gauge. The problem was put aside and
ascribed to some lack of understanding from us.
Recently however the same type of contributions were considered in
the Matrix String theory model of Ref. \cite{motl,dvv} where they
give origin to string configurations of different lengths.
To be more specific let us consider the Matrix String theory action:
\begin{equation}
\label{sym}
S = {1\over 2\pi } \int d\sigma d\tau~  \tr\left((D_\mu X^M)^2 + \theta^T 
\Dslash \theta + g_s^2 F_{\mu\nu}^2 - {1\over g_s^2}[X^M,X^S]^2 +
{1\over g_s} \theta^T\gamma_M [X^M,\theta]\right)
\end{equation}
where the fields $X^M$ ( $M=1,2,...8$) are $N\times N$ hermitian matrices,
as are the 8 fermionic fields $\theta^\a_L$ and $\theta^{\adot}_R$.
The two dimensional world sheet is an infinite cylinder parametrized by 
coordinates $(\sigma,\tau)$, with $\sigma$ between $0$ and $2\pi$.
This action can be obtained from the ten dimensional super Yang-Mills theory
by dimensional reduction and it features the same set of fields as 
Green-Schwarz action for type II superstring, except that here the fields are
matrices. In the limit where the string coupling constant $g_s$ goes to
zero the fields $X$ and $\theta$ will commute and can be simultaneously diagonalized.
The eigenvalues $x^M_i$ ($i=1,2,...,N$)  of $X^M$ can be identified with string coordinates
which describe the world sheets of a gas of $N$ Green-Schwarz light-cone strings.
The key point is that the eigenvalues $x^M_i$ can be interchanged as one goes
around the compact dimension $\sigma$: 
\begin{equation}
\label{perm}
x^M(\sigma+2 \pi) = P x^M(\sigma) P^{-1}~,
\end{equation} 
where $P$ is an element of $S_N$. In conclusion the fields $x^M_i(\sigma)$ 
take value on the orbifold space $S^N {\bf R}^8$, with strings of different
lengths associated to the cycles of $P$, and the corresponding Hilbert space
consists of twisted sectors in correspondence with the conjugacy classes of 
$S_N$.
\section{YM2 in the unitary gauge}
The same twisted sectors appear naturally in YM2 in the unitary gauge,
where the field strength $F$ is diagonal. 
Consider the partition function of YM2 in the
first order formalism on a general Riemann surface $\Sigma_g$ of genus $g$:
\begin{equation}
Z(\Sigma_g,t)=\int [dA][dF]\exp\left\{-\frac{t}{2}\tr \int_{\Sigma_g}
d\mu\, F^2+ \ii\, \tr \int_{\Sigma_g} f(A) F \right\}~,
\label{partriemann}
\end{equation}
where $d \mu$ is the volume form on $\Sigma_g$ and
$f(A)= d A -\ii A\wedge A$.
It is always possible, at least locally, to find a gauge 
transformation $g$ that diagonalizes $F$:
\begin{equation}
g^{-1} F g = \diag(\lambda)~.\label{Pgaugefix}
\end{equation}
However $g$ is not unique: if $g$ diagonalizes $F$, so does any gauge 
transformation of the form $g P$, with $P\in S_{N}$: in general, 
there are $N!$ Gribov copies of the gauge--fixed field strength $F$.
As in the case of Matrix string theory, the twisted sectors
appear because as  we go around a homotopically non trivial loop on 
$\Sigma_g$, the eigenvalues can be interchanged, namely we can go 
from one Gribov copy to another. 
Consider now the case where $\Sigma_g$ is a torus parametrized by coordinates
$(\sigma,\tau)$ both ranging from $0$ and $2 \pi$.
The twisted sectors are labelled by the pair of permutations $P$ and
$Q$ associated to the two homotopically non trivial loops, more 
precisely by the boundary conditions
\begin{eqnarray}
\lambda_i(\tau+2\pi, x)&=&\lambda_{P(i)}(\tau,x)~,\nonumber\\
\lambda_i(\tau, x+2\pi)&=&\lambda_{Q(i)}(\tau,x)~,
\label{bcon}
\end{eqnarray}
where consistency requires $P$ and $Q$ to commute:
$PQ=QP$.
\par
Pairs of commuting permutations also define $N$-coverings of 
the torus without branch points, the $N$ sheets of the covering at
each point of the target space being labelled by one eigenvalue of $F$. 
This argument is easily 
generalized to higher genus Riemann surfaces, and one can say in
conclusion that twisted sectors are in correspondence with the 
inequivalent $N$-coverings of $\Sigma_g$ in absence of branch points.
However if the genus is  greater than one the  quantization in
the unitary gauge leads to divergences whose regularization, according to
Ref. \cite{bt}, would amount to set to zero all twisted sectors.
The argument is as follows: the BRST invariant action in the unitary gauge
consists of two terms, the first, denoted by  $S_{\rm Cartan}$, 
depends only on the diagonal components of the gauge fields and is just 
the gauge action in the first order formalism for the residual 
$U(1)^N$ gauge invariance:
\begin{equation}
\label{scartan}
S_{\rm Cartan}= \int_{\Sigma_g}  \sum_{i=1}^N 
\left[\frac{t}{2}\lambda_i^2 d\mu -
\ii \lambda_i d A^{(i)} \right]~,
\end{equation}
where $A^{(i)}$ is the $i$-th diagonal term of the matrix form $A$.
The second term, named $ S_{\rm off-diag}$, contains the ghost and 
anti-ghost fields and  the non--diagonal components of $A$:
\begin{equation}
S_{\rm off-diag} =  \int_{\Sigma_g} d\mu \sum_{i>j}  (\lambda_i -
\lambda_j)\left[\hat{A}_0^{ij}\hat{ A}_1^{ji} -\hat{A}_1^{ij}\hat{A}_0^{ji} 
+ \ii ( c^{ij} \bar{c}^{ji}+ \bar{c}^{ij} c^{ji}) \right]~,
\label{offdiag}
\end{equation}
where $ \hat{A}_{a}^{ij} = E_a^{\mu} A_{\mu}^{ij} $ and $E_a^{\mu}$ denotes  
the inverse of the two dimensional vierbein.
$ S_{\rm off-diag}$  has a fermionic symmetry, which exchanges gauge and ghost 
fields; hence one would expect the contributions to the partition functions 
from the ghost fields and the non--diagonal part of the gauge fields to 
cancel exactly.
\par
However, this ``supersymmetry'' {\it is} in general broken by an anomaly
in the functional measure, due to the fact that on a curved surface the 
number of degrees of freedom of a one-form (like $A_{\mu}$) and of two zero 
forms (like $c$ and $\bar{c}$) do not match exactly. In fact the corresponding
functional integral has been calculated exactly in \cite{bt} and it is given 
by:
\begin{equation}
\int \prod_{i>j} [dc^{ij}][d\bar{c}^{ij}][dA_{\mu}^{ij}] e^{-S_{\rm off-diag}}
= \exp \left[{1 \over 8 \pi} \int_{\Sigma_g} R \sum_{i>j} \log (\lambda_i -
\lambda_j) \right]~,
\label{anomaly}
\end{equation}
where $R$ is the curvature scalar: only on flat Riemann surfaces like 
the torus or the cylinder the symmetry is preserved at the quantum 
level.
\par
Following \cite{bt} one finds, by gauge fixing the residual U$(1)^N$
invariance, that in the end only configurations where the eigenvalues
$\lambda_i$ are constant and equal to integers $n_i$ contribute. So the 
r.h.s. of (\ref{anomaly}) gives the standard dependence of YM2 partition
function from the genus, while the dependence from $t$ is given by the 
U$(1)^N$ action $S_{\rm Cartan}$ thus reproducing the well known partition 
function of YM2 on a Riemann surface \cite{ru,bt1,wit}:
\begin{equation}
Z(\Sigma_g,t) = \sum_{\{n_i\}} \frac{1}{\prod_{i>j} (n_i-n_j)^{2g-2}}
\esp{-2\pi^2 t \sum_i n_i^2}~.
\label{part}
\end{equation}
However in the present derivation nothing forbids $n_i=n_j$ for $i \neq j$,
and such terms, that we shall  name non-regular terms  following \cite{bt},
are divergent for $g>1$. Notice that non-regular terms always appear in 
the twisted sectors, where at least two sheets of the $N$-covering are 
connected by going round some non contractible loop. Therefore the 
corresponding eigenvalues are  given by the same integer.
The regularization  proposed in \cite{bt} is done by adding
mass terms to the non diagonal part of the gauge fields; in this way the
non-regular terms vanish due to the ghost contribution which is proportional
to $(n_i - n_j)^2$, while the contribution from the gauge fields is now finite
and proportional to $(n_i - n_j -m_{ij})^{-2g}$. The limit 
$m_{ij} \rightarrow 0$ is performed at the end.
This regularization scheme, while preserving the U$(1)^N$ gauge
symmetry, violates the original BRST invariance as well as the fermionic 
symmetry between gauge fields and ghosts discussed above.
A fully  consistent treatment of the non-regular terms in the
unitary gauge for $g>1$ is indeed lacking, and in our opinion 
this is a problem worth looking into in the future. The case $g=1$ is
special: non regular terms are finite, the fermionic symmetry is
anomaly free and hence there is no need of regularization.
Therefore we shall write the partition function as a sum over all sectors
labelled by commuting pairs of permutations $(P,Q)$, thus including 
non regular terms.
We will find that the result does not coincide with the standard partition
function \cite{ru,bt1}, which can be reproduced only by limiting the sum to 
the subset of sectors of the type $(P,1)$. This choice however is not 
invariant under modular transformations on the torus. 
\section{Free energy and partition function}
\label{secfree}
The twisted sectors can also be labelled, as discussed earlier on,
by the $N$-fold covers without branch points of the torus. 
In order to sum over all sectors it is convenient on one hand to work
in the grand canonical formalism, in which $N$ is not fixed, and introduce 
the grand-canonical partition function $Z(t,q)$ and the corresponding free 
energy $F(t,q)$:
\begin{equation}
\label{mondo1}
Z(t,q)= {\rm e}^{ F(t,q)} = \sum_N Z_N(t) q^N~.
\end{equation}
On the other hand it is convenient to work directly on the free energy 
$F(t,q)$, that receives contributions only from the connected coverings. 
The computation of the free energy entails two aspects: an ``entropic'' one, 
i.e. the counting of the inequivalent connected coverings, and the 
determination of the Boltzmann factor that the functional integral 
(\ref{partriemann}) implies for each covering. 
The counting of $N$-coverings of the torus without branch points, namely 
of the $N$-fold maps of a world sheet torus into the target torus, has 
already been discussed in the literature (see for instance \cite{ggt,ksw} ) 
and its free energy is given by
\begin{equation}
\label{torino1}
F_{\rm cov} = \sum_N \sum_{r|N} {1\over r} q^N = 
-\sum_{k=1}^\infty \log(1 - q^k)~
\end{equation}
where $r|N$ means that the sum is extended over the divisors $r$ of $N$.
The coefficient of $q^N$ in Eq. (\ref{torino1}), namely 
$\sum_{r|N} {1\over r} $,  
enumerates the {\em connected} $N$-coverings of the torus.
This result can be derived as follows: let the periods of the target space 
torus be $\vec\pi^{\rm tar}_1 = 2\pi$ 
and $\vec\pi^{\rm tar}_2 =2\pi {\rm i}$, the most general connected 
$N$-covering is then a torus of area  $4\pi^2 N$ whose periods are given by
\begin{equation}
\label{mondo4}
\vec\pi^{\rm ws}_1 = 2\pi k ~; ~~~~~~~~
\vec\pi^{\rm ws}_2 = 2\pi s + {\rm i}\, 2\pi r ~,
\end{equation}
with $kr =N$ and $s=0,1,\dots ,k-1$. 
\begin{figure}
\begin{center}
\null\hskip 1pt
\epsfxsize 9cm
\epsffile{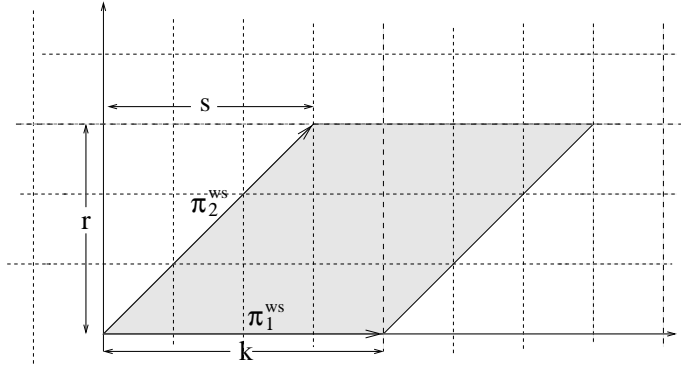}
\end{center}
\caption{A world-sheet torus in the case $N=12$, $k=4$, $s=3$.}
\end{figure}
An example with $N=12$, $k=4$ and $s=3$ is given in Figure 1.
There are $\sum_{k|N} k = N \sum_{r|N} 1/r$ choices of the integers $k$,$r$ 
and $s$ that satisfy these conditions, but one has to divide by the symmetry 
factor $N$ to account for the fact that coverings corresponding to different 
labelings of the sheets (i.e. of the eigenvalues) have to be identified. 
Notice that the world sheet torus has  a modular parameter 
$\tau = s/k + {\rm i}\, r/k = s/k + {\rm i} N/k^2$ 
and  that summing over these tori
with the weight $1/r = k/N$ is the 
discrete version of integrating over the modular parameter $\tau$ with the
usual modular-invariant measure. In fact from ${\rm Im}\tau=N/k^2$ and $N$
fixed we have $\int d{\rm Im}\tau/({\rm Im}\tau)^2 \to
\sum_{k|N}{k\over N}$.
\par
It is easy to see that in terms of the field $\lambda$, the partition
function (\ref{scartan}) becomes the partition function of a U$(1)$ theory 
on the world sheet torus. This partition function depends only on the area; 
it is well-known to be $\theta_3(0|{\rm i}\,2\pi N t)$ 
$=\sum_{n=-\infty}^\infty \exp (-2\pi^2 N t n^2)$. This is the Boltzmann weight
to  be associated to the connected coverings of degree $N$.
\par
We are now in the position of writing down the free energy, and thus
automatically the grand-canonical partition function:
\begin{eqnarray}
\label{torino2}
F^{\pm}(t,q) & = & \pm \sum_N \sum_{r|N} {1\over r} \sum_{n=-\infty}^\infty 
{\rm e}^{-2\pi^2 t N  n^2}\, q^N =
\mp \sum_{n=-\infty}^\infty \sum_{k=1}^\infty 
\log(1 -  {\rm e}^{-2\pi^2 t k n^2} q^k)~, \nonumber
\\
Z^\pm(t,q) & = &  
\prod_{n=-\infty}^\infty \prod_{k=1}^\infty (1 - {\rm e}^{-2\pi^2 k t n^2}
q^k)^{\mp 1}~.
\end{eqnarray}
\par
We have allowed in front of the free energy $F(t,q)$ a  sign ambiguity, which 
corresponds to different choices of the a priori undetermined weights with 
which the contributions from the different twisted  sectors are added.
It is clear from (\ref{torino2}) that the plus sign leads to partition 
function $Z^+$ of ``bosonic'' type (a ``state'' with fixed $k$ and $n$ may 
appear any number of times), while $Z^-$ is ``fermionic'' (the exclusion 
principle holds).
\par 
If we consider only the contribution of a subset of connected coverings, namely
those with $k=1$, $r=N$, that are associated to permutations of the type
$(P,1)$, we obtain the following  partition function:
\begin{equation}
\label{torino3}
{\cal Z}^\pm(t,q)  =   
\prod_{n=-\infty}^\infty (1 - {\rm e}^{-2\pi^2 t n^2}q)^{\mp 1}~.
\end{equation}
These expressions are known in the literature. 
${\cal Z}^-(t,q)$ is the grand-canonical expression that encodes, as shown in
\cite{us93}, the standard partition function for the U$(N)$ theory on the torus
\footnote{The coefficient of $q^N$ in the power series expansion of 
${\cal Z}^-(t,q)$ coincides, up to a sign and an overall normalization factor, 
with the standard partition function of  U$(N)$ only for odd values of $N$. 
For even values of $N$ the integers
$n_i$ are replaced by half-integers in the standard YM partition function.
This half-integer shift however can be re-absorbed by adding to the action
a term proportional to ${\rm tr} F$ (which is entirely in the U$(1)$ 
factor of U$(N)$). The expansion of ${\cal Z}^-(t,q)$ for even $N$ gives 
such modified theory. Notice that in the case of SU$(N)$  the problem does 
not arise and the sum over the $(P,1)$ sectors correctly reproduces the 
standard result for all values of $N$.} \cite{ru,bt1}. 
${\cal Z}^+$ reproduces the partition function obtained in \cite{het} by 
quantizing on the algebra rather than on the group.    
By comparing Eq.s (\ref{torino2}) and  (\ref{torino3}) we find
\begin{equation}
\label{torino4}
Z^\pm(t;q) = \prod_{k=1}^\infty {\cal Z}(kt;q^k)~.
\end{equation}
An expansion in powers of $q$ of both sides of Eq. \ref{torino4} leads
for the fermionic case to the following relation: 
\begin{equation}
\label{nuovo1}
(-1)^N Z^{-}_N(t) = \sum_{\{r_k\}}\delta(\sum_{k=1}^N k r_k - N) \,\prod_{k=1}^N 
(-1)^{r_k}{\cal Z}^{-}_{r_k}(kt)~,
\end{equation}
where $Z^{-}_N (t) $ is the U$(N)$ partition function including all sectors 
and ${\cal Z}^{-}_{r_k}(kt)$ is the standard U$(r_k)$ partition function 
(see however the discussion in the footnote). As an example we give the 
partition function for the case $N=3$:
\begin{equation}
\label{enne3}
Z^-_3(t) = {\cal Z}^{-}_3(t) - {\cal Z}^{-}_1(t){\cal Z}^{-}_1(2t)
+ {\cal Z}^{-}_1(3t)~.
\end{equation}
The extra terms at the r.h.s. of (\ref{enne3}) are related to the states with
$k>1$ and are not present in the conventional approach. 
\subsection{Modular invariance}
One of the new features of our approach is that the ensemble of twisted sectors
that contribute to the partition function is invariant under modular 
transformation on the cylinder, while the subset that reproduces the 
conventional result is not. In fact for instance a Dehn twist maps the 
sector labelled by $(1,P)$ into one labelled by $(P,P)$. 
As a result the conventional formulation is not modular invariant.
This does not show in the partition function, which depends only on the area,
but it should appear at the level of correlation function. For instance 
correlation functions of Wilson loops that are mapped into each other by
modular transformations  on the torus (like the two sets depicted 
in Figure 2) should coincide in our approach but not in the conventional 
formulation. Work is in progress to verify this point.
\begin{figure}[h]
\begin{center}
\null\hskip 1pt
\epsfxsize 8cm
\epsffile{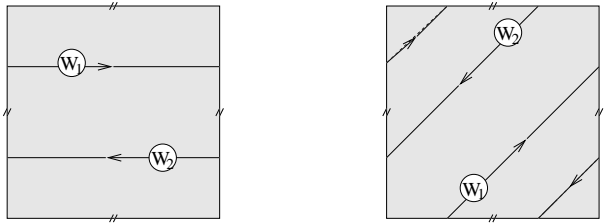}
\end{center}
\caption{Two sets of Wilson loops that are related by a Dehn twist.}
\end{figure}
\subsection{Quantization on the cylinder}
The partition function on a torus is often calculated by first considering
a cylinder with fixed holonomies at the edges and then by sewing the two ends
together. This involves identifying the two end holonomies up to a gauge 
transformation, namely identifying  their eigenvalues up to a permutation $P$.
Hence a sum over $P$ appears in the final result. It is clear that in this way
only the $(P,1)$ sectors are taken into account, and in order to consider also
sectors corresponding to non trivial commuting pairs $(P,Q)$, with $Q$ 
associated to the compact dimension of the cylinder, one has somehow to 
generalize the possible boundary conditions at the edges of the cylinder 
which in the conventional approach are just the U$(N)$ holonomies.
In order to do that let us choose the unitary gauge on the cylinder as in 
Eq. (\ref{Pgaugefix}), and consider a non trivial sector labelled by a 
permutation $Q$ that defines, according to the second of Eq.s (\ref{bcon}), 
the boundary conditions for the eigenvalues $\lambda_i(\tau,x)$ as we go 
round the compact dimension parametrized by the coordinate $x$.
We refer to the original paper \cite{us} for the details of the calculation and give 
here only the main result. 
This can be summarized by saying that for non trivial $Q$ there are as many
independent invariant angles in the holonomies as the number of cycles in $Q$.
More precisely, supposing that $Q$ has $r_k$ cycles of order $k$ 
the invariant angles $\theta_i$ of a Wilson loop winding round the compact 
dimension have the following structure:
\begin{equation}
\theta_{k,\alpha,n} = \theta_{k,\alpha}+\frac{2\pi\ii\,n}{k}~,
\label{holo}
\end{equation}
where we have made the replacement $i \rightarrow (k,\alpha,n)$ with 
$\alpha =1,\dots,r_k$ and $n=0,1,\dots,k-1$ to denote that $i$ is the 
$n$-th element of the $\alpha$-th cycle of order $k$.
The independent invariant angles are just the $\theta_{k,\alpha}$, and 
are associated to the cycles, the other eigenvalues within each cycle 
being spaced like the $k$-th roots of 1.
When sewing the cylinder the invariant angles are identified up to a 
permutation $P$ that preserves the cycle structure, namely
\begin{equation}
\label{ncyl2}
(k,\alpha,n)\stackrel{P}{\longrightarrow} (k,\pi_k(\alpha),n+ s(k,\alpha))~,
\end{equation}
where $\pi_k\in S_{r_k}$ is a permutations of the $r_k$ cycles of order $k$
and  $s(k,\alpha)$ is an integer shift ${\rm mod}\, k$. 
Eq. (\ref{ncyl2}) is equivalent to the statement that $P$ commutes with
$Q$, and hence it reproduces the by now familiar pattern of the twisted 
sectors.
{}From the previous discussion it is clear that the end states on the cylinder
are not parametrized by the U$(N)$ holonomies but rather by the holonomies
of ${\rm U}(r_1)\otimes {\rm U}(r_2)\otimes \dots $.
It is also clear that by considering just the U$(N)$ holonomies one is
automatically projecting on the trivial sector $Q=1$.
Although this projection appears the most natural thing to do in the 
framework of gauge theories, and it is so far also the only approach that 
we know how to implement on a lattice, it introduces an asymmetry between 
the two generators of the torus and ultimately breaks the modular invariance 
in the sense mentioned above.
\section{YM2 as a Matrix String Theory.}
The grand canonical partition function $Z^{\pm}(t,q)$ given in 
Eq. (\ref{torino2}) describes in the  large $t$ limit the coverings of the 
torus. In fact in that limit the Boltzmann factor given by the partition 
function of the U$(1)$ gauge theory tends to $1$ and we are left with the 
partition function that simply counts the homotopically distinct coverings
of the $g=1$ target space by a $G=1$ world-sheet: 
$Z^{\pm}(t,q) \stackrel{t \to \infty}{\rightarrow}  Z_{\rm cov}(q)$. 
So in this limit the twisted sectors define a string theory, exactly in the 
same way as they do in Matrix string theory. It can also be argued, and 
in the fermionic case it is obvious from the exclusion principle, that in 
this limit large values of $k$, namely long strings, are relevant.
Quite the opposite happens in the limit $t \to 0$, in fact one can see from 
the Poisson re-summation formula that in this limit the Boltzmann weight 
behaves like $1/\sqrt{t}$, and so the leading contributions to the 
partition function come from coverings that maximize the number of 
disconnected world sheets. So only states with $k=1$ survive in the 
$t \rightarrow 0$ limit, and the theory reduces to the conventional one 
without twisted sectors.
It is possible, but yet not proved, that  a phase transition separates  the 
two phases dominated respectively by long and short strings.
The small $t$ region is also the relevant one for the large $N$ limit of 
YM2 studied by Gross and Taylor in a series of papers \cite{ggt}. 
In fact the limit is done, following the original idea of 't Hooft, by 
taking $t = \tilde{t}/N$ and keeping $\tilde t$ fixed. Gross and 
Taylor proved that the partition function of $U(N)$ Yang-Mills theory on a
two dimensional Riemann surface $M_g$ of genus $g$ counts the homotopically 
distinct maps from a world-sheet $W_G$ of genus $G$ to $M_g$.
In particular if the target space is a torus and one considers only the leading term
in the large $N$ expansion, which means considering only world sheets of genus $1$,
one finds the partition function
\begin{equation}
\label{grosspf}
Z_{GT} = \prod_{k=1}^{\infty} (1 - {\rm e}^{-k \tilde t} )
\end{equation}
This is the same string partition function that is found from the 
twisted sectors in the large $t$ limit, with $q$ replaced by
${\rm e}^{- \tilde t}$. This coincidence is suggestive of some kind of
underlying duality in the theory.
However for this duality to exists beyond the rather trivial case of $W_{G=1}
\rightarrow M_{g=1}$ maps we should be able to extend the twisted sectors 
introduced in \cite{us} to include coverings with branch points, which 
correspond to higher genus world sheets, namely to the possibility for 
strings to split and join. These would be dual to the sub-leading terms in 
the $1/N$ expansion of \cite{ggt}.
However branch points correspond to points where the curvature of the 
world sheet becomes a delta function, leading through Eq. (\ref{anomaly}) 
to terms coupling different sheets of the coverings.
The problem here is of the same nature as the one that is encountered if one 
quantizes YM2 in the unitary gauge on a surface with $g>1$, namely the 
appearance of logarithmic interactions between different eigenvalues, 
that eventually lead to divergences. A better understanding of this point 
is then crucial also for a consistent formulation of YM2 as a matrix string 
theory on a torus, so it is not yet clear to us if such 
formulation exists also at the level of string interactions, or weather this 
can be done only by embedding YM2 in the more general framework of the 
Matrix String Theory given by (\ref{sym}), where it is known that string 
interactions can be consistently introduced \cite{bobone}.
In both cases the analysis developed in \cite{us} is relevant. 
In fact gauge degrees of freedom are crucial in Matrix String theory  
for the description of string interactions
as well as for the relation of non-trivial fluxes with D-brane charges
\cite{dvv}.  
Shortly after our paper appeared on the hep-th archive, Kostov and 
Vanhove \cite{kv} calculated the partition function of the Matrix String 
Theory of Eq. (\ref{sym}). They took advantage of the fact that the 
contributions to the partition function of the "matter fields" $X^M$ and 
$\theta$ cancel due to supersymmetry, so that in the end the only 
contributions come from the different topological sectors of YM2. 
In fact their result coincides with ours, except that they obtain our 
free energy rather than the partition function because the structure of 
the fermionic zero-modes in the matter sector effectively kills the 
contributions from disconnected world sheets. In \cite{kv}
the dimensional reduction of this free-energy to zero dimensions is shown to
correctly reproduce the partition function for the IKKT model \cite{ikkt},
namely $Z_{\rm IKKT}\propto \sum_{r|N}{1\over r^2}$. The latter \cite{gg}
is related to the moduli space of D-instantons or, by T-duality, to the 
counting of bound states of D0-branes.
\end{document}